\documentclass[aps,floatfix,prc,twocolumn,superscriptaddress,showpacs,
 amsmath,amssymb,
]{revtex4}
\usepackage{enumerate}
\usepackage{amsmath}
\usepackage{graphicx}% Include figure files
\usepackage{dcolumn}% Align table columns on decimal point
\usepackage{array}
\usepackage{bm}% bold math
\usepackage{tabularx}
\usepackage{siunitx}
\usepackage{booktabs}
\usepackage{blindtext}
\usepackage{hyperref}
\usepackage{adjustbox}
\usepackage{verbatim}
\usepackage{color}
\usepackage{subcaption}
\usepackage{natbib}
\begin{document}

\title{Propagation of Hauser-Feshbach uncertainty estimates to r-process nucleosynthesis: Benchmark of statistical property models for neutron rich nuclei far from stability.}% Force line breaks with \\

\author{S. Nikas}\altaffiliation{Present Address: Tecnische Universit\"{a}t Darmstadt and GSI Helmholtzzentrum f\"{u}r Schwerionenforschung, Darmstadt, Germany}
\affiliation{Department of Physics, Central Michigan University, Mount Pleasant, Michigan 48859, USA}
\affiliation{Joint Institute for Nuclear Astrophysics, Michigan State University, East Lansing, Michigan 48824, USA}
\author{G. Perdikakis}\email{perdi1g@cmich.edu}
\affiliation{Department of Physics, Central Michigan University, Mount Pleasant, Michigan 48859, USA}
\affiliation{Joint Institute for Nuclear Astrophysics, Michigan State University, East Lansing, Michigan 48824, USA}
\affiliation{National Superconducting Cyclotron Laboratory (NSCL), Michigan State University, East Lansing, Michigan 48824, USA}

\author{M. Beard}\altaffiliation{deceased}
\affiliation{Department of Physics, University of Notre Dame, Indiana 46556, USA}

\author{R. Surman}
\affiliation{Joint Institute for Nuclear Astrophysics, Michigan State University, East Lansing, Michigan 48824, USA}
\affiliation{Department of Physics, University of Notre Dame, Indiana 46556, USA}
\author{M. R. Mumpower}
\affiliation{Joint Institute for Nuclear Astrophysics, Michigan State University, East Lansing, Michigan 48824, USA}
\affiliation{Theoretical Division, Los Alamos National Laboratory, Los Alamos, New Mexico 87545, USA}
\author{P. Tsintari}
\affiliation{Department of Physics, Central Michigan University, Mount Pleasant, Michigan 48859, USA}
\affiliation{Joint Institute for Nuclear Astrophysics, Michigan State University, East Lansing, Michigan 48824, USA}
\date{\today}% It is always \today, today,

\begin{abstract}
\textbf{Background:} The multimessenger observations of the neutron star merger event GW170817 have re-energized the longstanding debate over the astrophysical origins of the most massive elements via the r-process nucleosynthesis.
%The discovery by LIGO of nucleosynthesis evidence in the merging event of two neutron stars and first wide application of multi-messenger astronomy created new opportunities in the quest to understand the origin of elements. 
A key aspect of r-process studies is the ability to compare astronomical observations to theoretical calculations of nucleosynthesis yields in a meaningful way. To perform realistic nucleosynthesis calculations, understanding the uncertainty inherent in microphysics details such as nuclear reaction rates is as essential as understanding uncertainties in the modeling of the astrophysical environment.  
\newline \textbf{Purpose:} In this work, we present an investigation of the uncertainty of neutron capture rate calculations using the Hauser-Feshbach model when they are extrapolated away from stability. This work aims to provide a quantitative measure of the dependability of Hauser Feshbach calculations when the models of statistical nuclear properties (level density and gamma-ray strength) that are tuned near or on stability are extrapolated to nuclei in an r-process network.
\newline \textbf{Methods:} 
We have selected from literature a number of level density and gamma-ray strength models that are appropriate for describing neutron-capture reaction cross sections and used them to calculate the neutron-capture rate of each nucleus participating in the network. All valid model combinations for temperatures between $10^{-4}$~GK and 10~GK were used. In each calculation, we extracted E1 gamma-ray strengths and level densities and observed how these statistical properties affect the theoretical reaction rates. The resulting neutron capture rates were sampled with the Monte Carlo technique and used in r-process nucleosynthesis network calculations to map the range of possible results for the r-process abundances.    
\newline \textbf{Results:} The results show that neutron capture rates calculated with the extrapolated models of statistical nuclear properties can vary by a couple of orders of magnitude between different calculations. Phenomenological models provide smoother results than semi-microscopic ones. They cannot, however, reproduce drastic changes in the nuclear structure such as shell closures. While the semi-microscopic models examined in this work do predict nuclear structure effects away from stability, it is not clear that these results are quantitatively accurate. %enough. 
The overall effect of the extrapolation uncertainty to the r-process nucleosynthesis yields has shown to be large enough to impede comparisons between observation and calculations.
\newline\textbf{Conclusions:} Microphysics details such as neutron capture rates affect in a significant way the outcome of nucleosynthesis calculations. The inherent uncertainty in extrapolations of the current Hauser-Feshbach theory away from stability presents a challenge to meaningful comparisons of the results of nucleosynthesis calculations with observations. Based on the results of this study it is suggested that progress in the development of better microscopic models of gamma strengths and level densities is urgently needed to improve the fidelity of r-process models.ns of the neutron star merger event GW170817 have re-energized the longstanding debate over the astrophysical origins of the most massive elements via the r-process nucleosynthesis. 
\end{abstract}

%Use showkeys class option if keyword
\maketitle

\section{Introduction}
\begin{figure}
    \includegraphics[width=1\columnwidth]{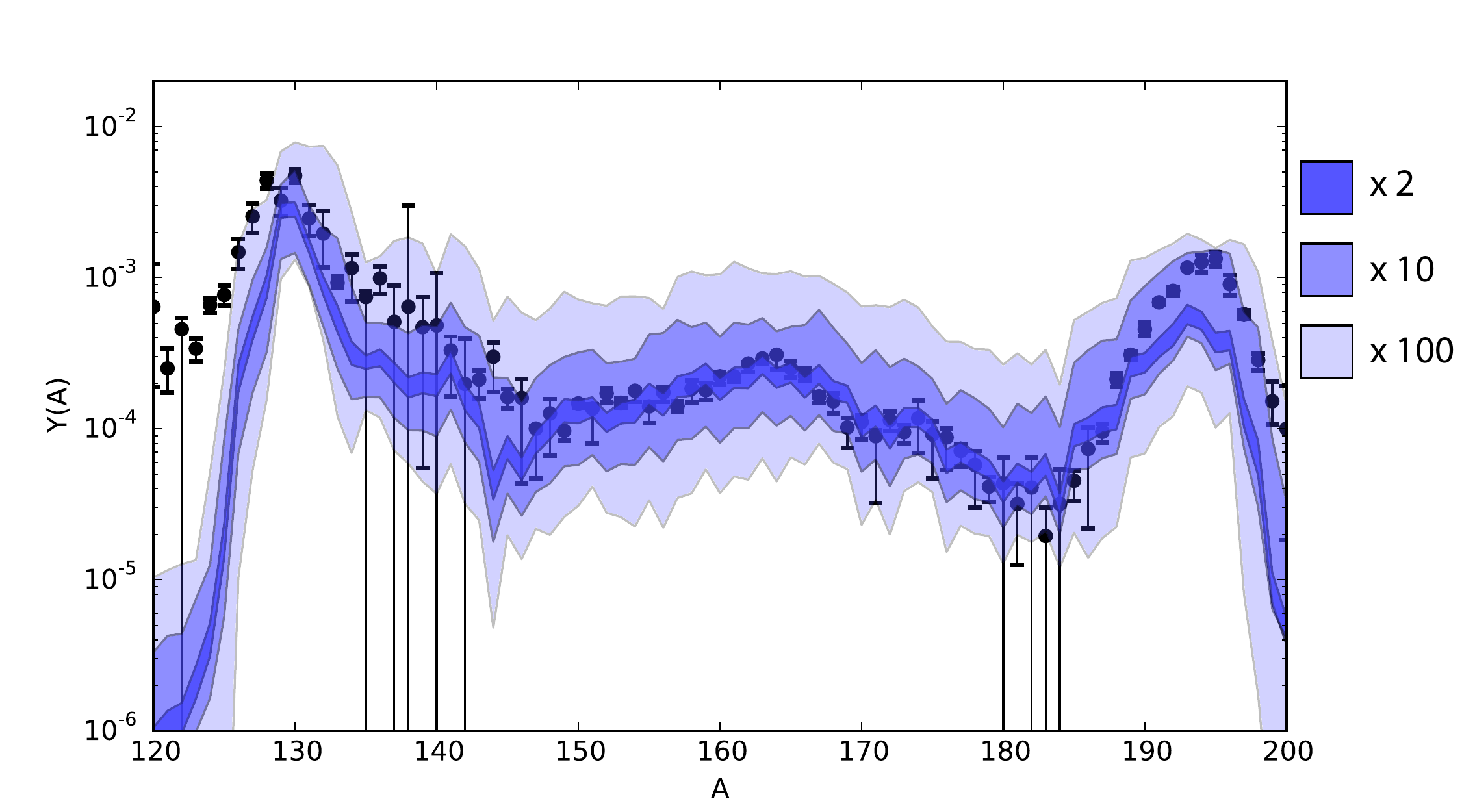}
	\caption{Monte Carlo study of the effect of reaction rate uncertainties to calculations of nucleosynthesis yields in an r-process example. The blue shaded area corresponds to different assumptions about the uncertainty of the neutron capture reaction rates. A factor of 10 uncertainty in neutron captures is enough to smear any fine details of the abundance pattern \cite{PhysRevLett.116.242502, mumpower2016impact}\label{fig:R-process_Impact2}}
\end{figure}

The scientific community has long pursued an answer to the question of the origin of elements. The seminal work of Margaret Burbidge et al. (commonly referred to as the B$ ^2 $FH paper) \cite{b2fh} in the fifties, defined our main ideas about nucleosynthesis in stars. Following that work, a whole field of nuclear astrophysics emerged, and we have now, after more than half a century, associated basic mechanisms with the synthesis of most of the elements in the periodic table \cite{Iliadis07a}. The discovery by LIGO of gravitational signature for merging neutron stars and the subsequent verification of nucleosynthesis evidence by various observations (see e.g., \cite{Hor19a} and references within), presents an excellent opportunity to understand the details of heavy element nucleosynthesis in such environments, as well as in other candidate sites. A critical aspect of this work is our ability to compare observation to theoretical calculations of nucleosynthesis yields. Beyond the description of the astrophysical environment, such calculations depend on the microphysics, including the contribution of nuclear physics uncertainties to nucleosynthesis predictions. \\
%One way to account for the uncertainty is by sensitivity studies. Mumpower and Surman et al. \cite{mumpower2016impact}, \cite{Sur14a} have used variations by constant factors of 2,10,and 100 to determine the effect of masses, $ \beta $-decay rates and neutron capture reaction rates to the yields. Using Monte-Carlo techniques, they have folded randomly these variation factors into nucleosynthesis network calculations and investigated the statistical significance of such variations and how it affects our ability to compare theoretical yields to observation. The result of one such study that we present
One way to quantify the impact of nuclear uncertainties on observables such as abundance patterns is through Monte Carlo rate variations. In the work of \cite{mumpower2016impact}, \cite{Sur14a}, the effects of mass, $\beta$-decay half-life, and neutron capture rate uncertainties were explored by folding random variations of these quantities, pulled from Gaussian distributions of constant width, into nucleosynthesis network calculations. The result of one such study presented in figure \ref{fig:R-process_Impact2}, is that a randomly assigned uncertainty of neutron capture rates of a factor of 10 is enough to smear characteristic features of theoretically calculated abundance patterns \cite{PhysRevLett.116.242502}. Such levels of uncertainty could impair our ability to perform meaningful comparisons between calculations of r-process yields and observation data. \\
A natural question connected to the results of
Monte Carlo rate variation studies is what uncertainty factor is reasonable to assume for calculated neutron capture reaction rates when the Hauser-Feshbach statistical framework is used with extrapolated statistical nuclear properties in order to describe nuclear reactions away from stability. 
Previous work (\cite{PhysRevLett.116.242502}, \cite{Denissenkov2018}, \cite{McKay2019a}) has addressed this question in limited regions of the nuclear chart using as an uncertainty estimate the variation in the reaction rates calculated by using different models of nuclear statistical properties. For neutron captures, the critical statistical properties are the nuclear level density (NLD) and the gamma-ray strength function ($\gamma$SF). In this work, we extend our previous approach to map with our uncertainty estimates the whole region of neutron-rich nuclei that may participate in r-process scenarios. We utilize the uncertainty estimates in a Monte-Carlo study to explore how they are propagated to nucleosynthesis. Furthermore, we investigate the origin of such uncertainties in the Hauser-Feshbach framework and connect it to the physics description of the major nuclear statistical properties away from stability. We explore how the disparity between various models of these quantities evolves along isotopic lines fueled by the lack of experimental constraints. We make model suggestions for future calculations as well as requests for experimental investigations that could improve the current state of the theory.\\
The manuscript is organized in the following way: In section \ref{CalculationsHF} we describe the details of the reaction rate calculations we performed. In section \ref{MonteCarlo} we provide the details of the Monte-Carlo calculations of the nucleosynthesis yields based on our reaction rate calculations, and in section \ref{Results}, we present and discuss the results of this study and how it informs reaction rate and nucleosynthesis calculations. Last, in section \ref{Conclusion}, we summarize this work and provide some insights regarding future theoretical and experimental developments.

%In this manuscript we compare the behavior of various theoretical models of statistical nuclear properties away from stability and as a function of stellar temperature, we define uncertainty factors based on the maximum calculated variations of reaction rates, and then use the range of calculated reaction rates for each isotope in a Monte-Carlo variation study of nucleosynthesis yields under various r-process scenarios.
\section{Description of Hauser-Feshbach calculations}\label{CalculationsHF}
Theoretically predicted neutron capture rates for the r-process are typically calculated using the Hauser--Feshbach model \cite{PhysRev.87.366}. This theoretical model describes the decay of highly excited nuclei characterized by a large number of levels per MeV of excitation energy, provided that reasonable descriptions of the appropriate nuclear statistical properties and nuclear potentials are available. Two types of uncertainty have to be evaluated for these calculations: (1) the intrinsic uncertainty due to the fact that the statistical model is expected to only reproduce an average behavior of highly excited nuclei, and (2) The extrapolation uncertainty stemming from the fact that the ingredients of the calculation (level densities, potentials,  gamma-ray strengths, etc.) are not necessarily describing the nuclei accurately, especially far away from stability where little or no experimental data exists.\\ 
The intrinsic uncertainty can be reduced by constraining the statistical properties and nuclear potentials with experimental data on reaction rates.  For neutron capture reactions, such measurements are only possible for stable or long-lived isotopes. Beard et al. \cite{Beard2014} has evaluated the intrinsic uncertainty of Hauser-Feshbach predictions by comparing predicted reaction rates with experimental data and found an uncertainty that is of the order of a factor of 2.\\
Evaluating extrapolation uncertainty is a more challenging task. In the absence of experimental data to compare for unstable nuclei, we are confronted with several
potentially sound models that can be used on an equal footing judging from their performance in the description of highly excited stable nuclei. To our best knowledge, the theoretical uncertainty in such a situation is given by the range of calculated values. This is the approach followed in \cite{PhysRevLett.116.242502} and \cite{Denissenkov2018} and that we apply in this work to a larger body of nuclei extending further away from stability. 

We use the statistical model code TALYS \cite{Kon08a} to calculate the reaction rates. For consistency with earlier r-process studies since many of the nuclei studied here have no associated experimental mass data, we calculated the Q-values for all reactions from the tabulated masses of P. Moller et al. \cite{Moller95}. We calculated the transmission coefficients for the formation of the compound nucleus from a spherical neutron-nucleus optical potential using the Koning-Delaroche globally fitted parameters \cite{Kon03a}. Calculations using the semi-microscopic approach of Bauge et al. \cite{Bau98a}-- based itself on the Jekeune-Lejeune-Mahaux \cite{Jeu77a} Nuclear Matter approach-- yielded less than 50\% difference in the reaction rate in comparison. Generally, variations in the capture rate roughly reaching a factor of 2 can be expected with a reasonable adjustment of parameters for nuclei on or near stability \cite{Mumpower2017} and higher off stability \cite{Jones2019}. It was, therefore, considered sufficient to use the Koning-Delaroche optical potential \cite{Kon03a} for all calculations. \\
The excited states for the compound and residual nuclei were described using available experimental spectroscopic information up to the tenth excitation energy level of any nucleus. Above the tenth level or otherwise above the last experimentally known level (e.g., for more exotic species) level density models chosen among the ones available with the TALYS code were used, as shown in table \ref{table:NLD_GSF_models}. We adopted five of the six nuclear level density models available in TALYS. The temperature-dependent Hartree Fock-Bogolyubov level densities using the Gogny force based on Hilaire's combinatorial tables \cite{Hil12a} were found during test calculations to produce unphysically strong odd-even effects for some neutron capture reaction rates away from stability. This could be related to systematic disagreements with experimental data mentioned by Hilaire et al. in \cite{Hil12a} for odd-odd and odd-A nuclei. Furthermore, this model is often unable to reproduce the almost exponential behavior observed experimentally for the level density (see \cite{Hil12a, Voi09a, Gut13a}). This possibly unphysical behavior warrants further investigation, and hence the corresponding level density model was deemed not suitable for this work.\\
\begin{table*}[h]
    \centering
    \caption{List of models used to describe the NLD and the $\gamma$SF in the Hauser-Feshbach calculations of this work. The first three NLD models are phenomenological. The two lower ones are semi-microscopic. For the $ \gamma $SF (right column), the two Hartree-Fock models are semi-microscopic, while the rest are phenomenological.}
\resizebox{\textwidth}{!}{
                \begin{tabular}{{cc}}
                    \hline\hline
            Nuclear Level Density models&$\gamma$ ray Strength Function models\\
            \hline\hline
            Constant Temperature and Fermi Gas \cite{Dil73a}&Kopecky-Uhl generalized Lorentzian \cite{Kop90a}\\
            Back-shifted Fermi Gas \cite{Dil73a},\cite{Gil65a}&Hartree-Fock BCS $+$ QRPA \cite{Gor02a} \\
            Generalized Super fluid \cite{Ign79a}, \cite{Ign93a}&Hartree-Fock-Bogolyubov $+$ QRPA \cite{Gor04a}\\
            Hartree Fock using Skyrme force \cite{Gor01a}&Modified Lorentzian \cite{Gor98a}\\
            Hartree-Fock-Bogoliubov and combinatorial \cite{MLD_Gor-Hil}&\\
            \hline
        \end{tabular} 
        \label{table:NLD_GSF_models}
}
\end{table*} 
The gamma decay of the compound nucleus was treated in the usual approach (\cite{Bartholomew1973}), assuming the validity of the Brink hypothesis and using M1 strength normalized to the E1 strength according to the Reference Input Parameter Library -2 (RIPL-2) prescription. The E1 strength was described using the phenomenological and semi-microscopic models available in the TALYS code. We limited the current study to the various modeling approaches for the E1 strength and chose not to include more exotic physics that have been suggested and which could have a large effect on the neutron capture rate (for example, we did not consider the possibility of M1 upbend). Out of the 5 available E1 $ \gamma $-ray strength function ($\gamma$SF) parameterizations, we adopted four. The  Brink-Axel single Lorentzian formula \cite{Bri57a, Axe62a} is known to exhibit a cut-off at lower $\gamma$ ray energies at the limit of E$_{\gamma}\rightarrow 0$ that has been shown not to agree with experimental data for neutron capture reactions (see e.g., \cite{Kop90a}). In general, this model is known to consistently overestimate average and total radiative widths as well as experimental neutron-capture cross sections for stable nuclei (for example, \cite{Kop90a, Cap09a}). Therefore it was also not used in any of the results shown in this work.\\
One last comment is in order to complete our perspective on reaction rate calculations away from stability. Particular note has to be taken of the fact that even along the line of beta stability, phenomenological models of Gamma-ray strength need to be renormalized to reproduce the experimentally observed strengths. This inherent limitation of the predictive value of these models is mitigated typically by the use of tabulated renormalization values that are A-dependent and valid along or very near the stability line. Naturally, such a table does not exist for the isotopes of interest to r-process studies, and this adds to the uncertainty of Hauser-Feshbach calculations of neutron capture rates along the r-process path.
\section{Monte-Carlo Analysis} \label{MonteCarlo}
We can appreciate the effect of nuclear reaction rate uncertainties on nucleosynthesis quantitatively using the Monte Carlo statistical technique. The approach and its implementation are described in detail in \cite{mumpower2016impact}. Each reaction rate that enters a nucleosynthesis network calculation is sampled from a distribution of theoretically predicted values for that reaction rate. This process we repeat for a large number of nucleosynthesis calculations assuming the same astrophysical environment. In each step, we record the nucleosynthesis yields. After the large set of calculations has been completed a range of values for each isotopic abundance has been calculated. The distribution of yield values corresponds to the distribution of reaction rate values that we initially input into the calculation.

The sampling of the reaction rates is done using a flat distribution of the various models, as it is statistically acceptable to do in cases where the distribution of values is unknown. As described above, away from stability, in the absence of experimental data, we have no grounds to assume that one of the models used is more accurate than the others. The uncertainty estimate for the yields arises from the variation achieved by the ensemble of nucleosynthesis calculations.
\section{Results}\label{Results}
\subsection{Hauser-Feshbach Extrapolation Uncertainties}\label{section:HauserFeshbach}
To better visualize the range of results possible between different Hauser-Feshbach extrapolations for a given temperature $T$, we use the logarithm of the ratio between the highest and lowest rates calculated for each reaction
\begin{equation}
\alpha=\log\frac{R_{max}}{R_{min}}
\end{equation} where $R_{max}$ and $R_{min}$ correspond to the calculations with the highest and lowest reaction rate value respectively for a given number of representative calculations $R_i $. A value of 0 for $ \alpha $ corresponds to an agreement between the various extrapolated calculations, while a value of 1 corresponds to a factor of 10 (one order of magnitude) variation between the extrapolations. The $ \alpha $ index is, in some sense, indicative of the uncertainty inherent in estimates of reaction rates using the Hauser-Feshbach framework at the various areas of the nuclear chart. We use the quantity $ \alpha $
to map the range of results between the various extrapolation-based neutron capture rates for neutron-rich nuclei with an atomic number between 10 and 100. Naturally, due to the Hauser-Feshbach theory's limitations, we do not consider the calculated rates to be quantitatively accurate for light nuclei with very low numbers of levels. Similarly, we have not performed any optimization of our calculations for fission, and hence the same disclaimer holds for fissioning nuclei. These limitations we feel do not subtract from the general aim of this work, which is to benchmark the effect of statistical property modeling to neutron capture rates. The resulting map overlayed on the table of isotopes for a stellar temperature of 1 GK is shown in figure \ref{fig:Ratio1GKnew} (See Supplemental Material at [URL will be inserted by publisher] for the table of corresponding $\alpha$ values). This temperature is chosen since at higher temperatures it is likely that the r-process proceeds in an ($n$,$\gamma$)-($\gamma$,$n$) equilibrium that renders the nucleosynthesis largely insensitive to individual capture rates; only after equilibrium fails do the nucleosynthetic yields become dependent on the rate of neutron captures.

\begin{figure*}[h]
    \includegraphics[width=\textwidth]{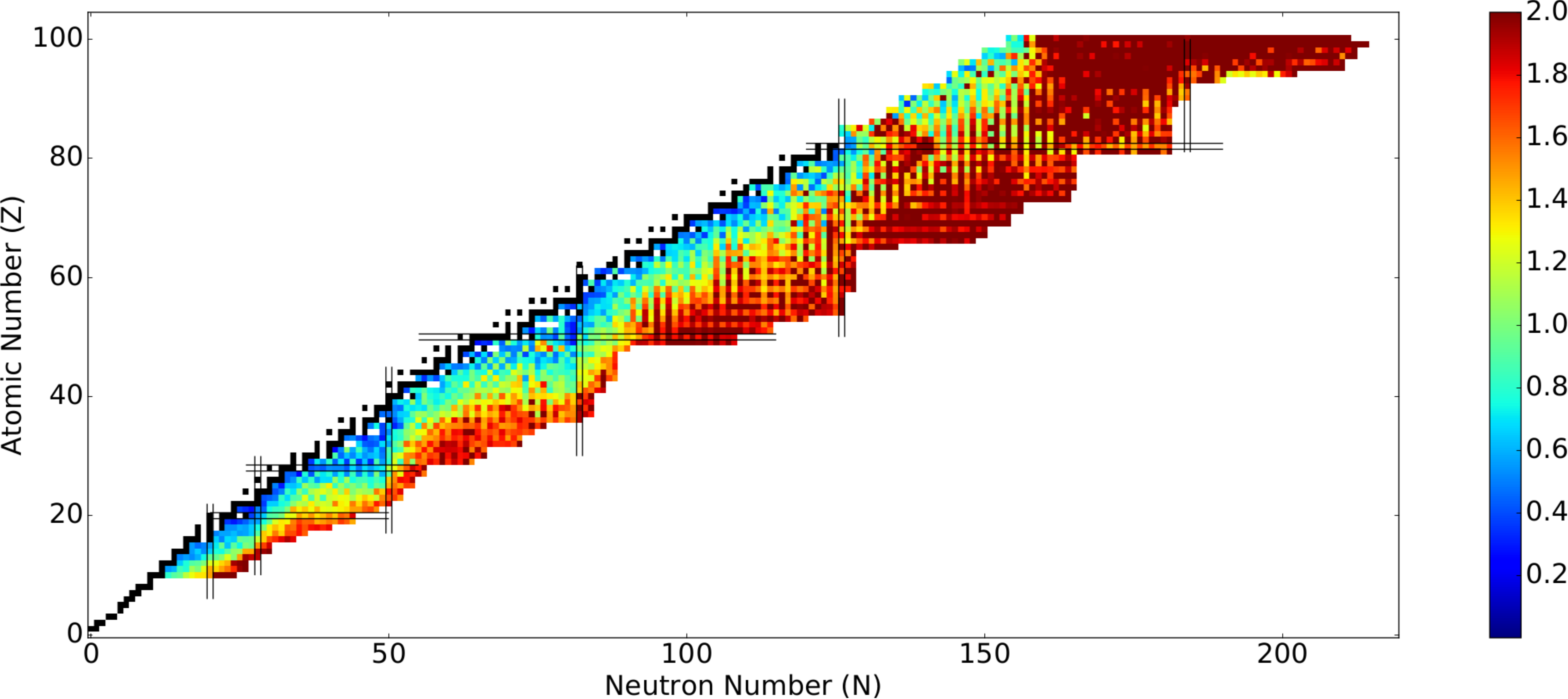}
    \caption{Map of agreement $ \alpha $ as a function of proton and neutron number at the neutron-rich side of the table of isotopes. A value of 0 for $ \alpha $ corresponds to an agreement between the various extrapolated calculations, while a value of 1 corresponds to a factor of 10 (one order of magnitude) variation between the reaction rate calculations. The $ \alpha $ factor serves as a benchmark of the effect of the modeling of statistical properties only. In this work, we have not attempted to amend the usual limitations of Hauser-Feshbach calculations for nuclei with very few excited levels or to optimize the calculation of fission. }\label{fig:Ratio1GKnew}
\end{figure*}
We observe that $\alpha $ ranges from 0 to at least 2 in figure \ref{fig:Ratio1GKnew}, indicating that the various extrapolations produce rates that differ by up to a few orders of magnitude in the worst case. With few exceptions, the increase in the neutron richness corresponds to decreasing agreement between rates determined with the HF model sets shown in Table I. 

Questions on possible correlations of uncertainty calculations arise naturally from such a comprehensive study. Identifying such correlations and characterizing them could help further constrain Hauser-Feshbach estimated reaction rates. In this work, we examine if there is a correlation between high $\alpha$ values and the estimates of specific statistical properties by the various models. We also check whether some statistical properties are more robustly estimated, hence better suited to extrapolations and if any disagreement between calculations is due to numerical issues or because of predicted changes in nuclear structure. We study to what extent the extrapolated reaction rates smoothly diverge along isotopic chains or if there are points of rapid deterioration of agreement between calculations indicated by high $ \alpha $ values. Last, we briefly examine the correlation of extrapolated reaction rates with stellar temperature. One more caveat related to the nature of these calculations exists and has to be noted when examining figure \ref{fig:Ratio1GKnew} for isotopes near the dripline; No model of direct capture is included in the calculations although it is expected to be important as the dripline is approached. However, we do include reaction channels with multiple neutrons in the exit channel (e.g. n,2n) that are expected to increase in importance as the neutron separation energy decreases.  

For the r-process, it is also interesting to investigate how the models behave when estimating neutron capture reaction rates on isotopes important for the creation of the r-process abundance pattern. In this work, we use neutron capture rates on isotopes of Eu and Ga to benchmark the behavior of the theory in two different mass regions that are of interest to the r-process (as suggested in \cite{mumpower2016impact}). We present our benchmark results for the level density and E(1) gamma-ray strength functions in subsections \ref{subsection_NLD_n_rich} and \ref{subsection_GSF_n_rich}. We show the global effects to reaction rates by isotope using europium and gallium isotopes as examples in subsection \ref{subsection_Effect_on_rates}. We dedicate a subsection (\ref{subsection_special}) to focus on the results of our calculations for two isotopes of particular interest for the r-process, $ ^{165} $Eu and $ ^{81} $Ga. The section ends with the results of the Monte-Carlo propagation of our estimated uncertainties to r-process nucleosynthesis at a few environments of current interest (subsection \ref{subsection_MC_nucleosynthesis}).

\subsection{Results for Nuclear Level Density}\label{subsection_NLD_n_rich}
Following the overall trend of neutron separation energy, the nuclear level density should decrease with an increasing number of neutrons in the nucleus. The particularities of each nucleus' nuclear structure modulate this overall trend, creating local features in the level density. There is a significant difference in the way the phenomenological and the microscopic models address these local effects. 

The phenomenological models use an analytical formula based on the liquid drop mass model parametrization to calculate contributions from nuclear structure effects. The formula introduces an energy shift in the calculation of the energy-dependent average level density parameter, $a$. The magnitude of this energy shift (the so-called shell correction) depends on whether the nucleus is even-even, odd, or odd-odd, and on the particular mass model used to calculate the neutron separation energy. 

The semi-microscopic models that we have used in this work employ Hartree-Fock methods to calculate the level density. These microscopic calculations dictate the local nuclear-structure-related variations of the level density, and hence it is expected that these models will reflect the predictions of the Hartree-Fock calculations for the nuclear structure of nuclei away from stability. 

\begin{figure}[h]
    \includegraphics[width=\columnwidth]{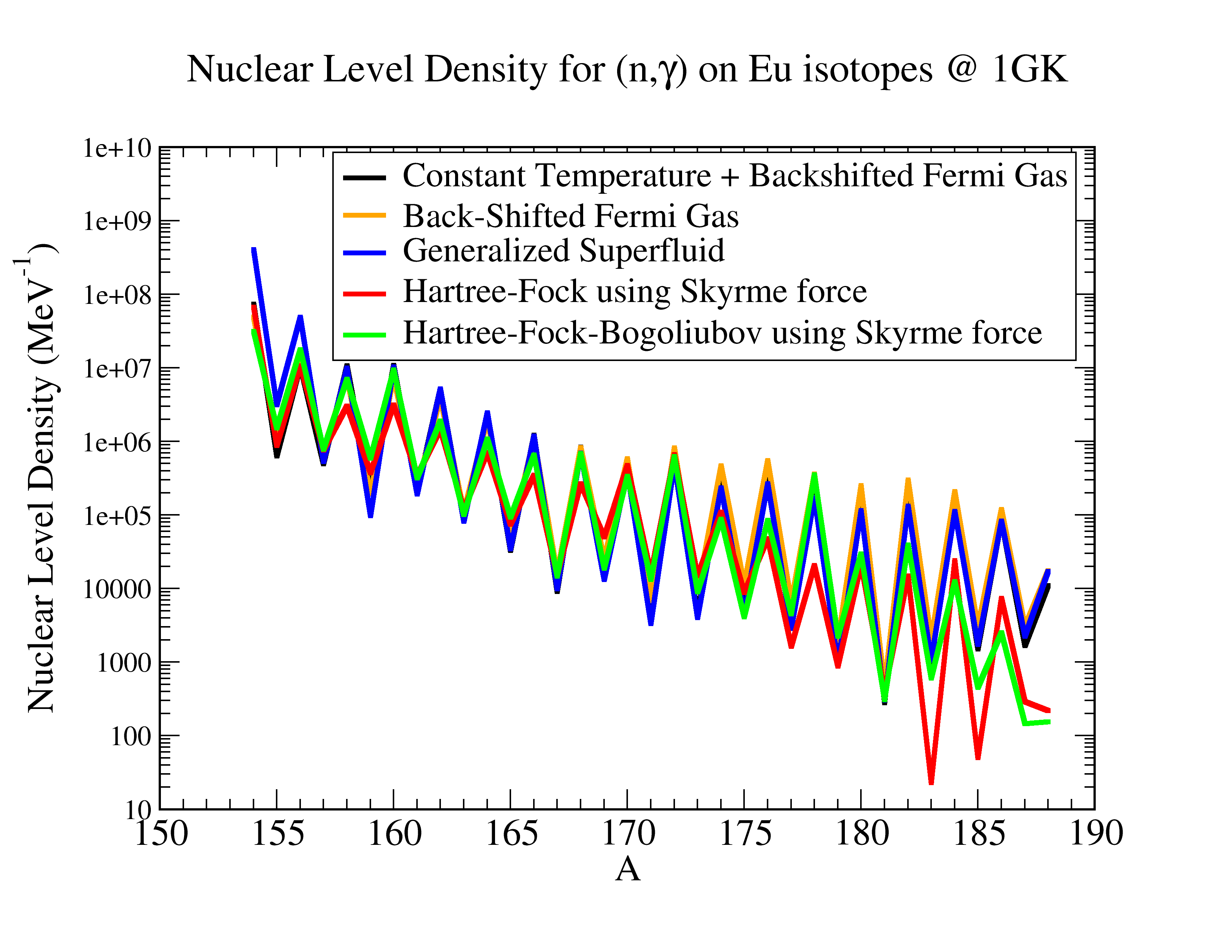}
    \caption{Plot of calculated nuclear level density in neutron capture using various models, for neutron-rich isotopes of europium. We plot the level density for the compound nucleus created by neutron capture at a stellar temperature of 1~GK. The horizontal axis corresponds to the mass number of the target in the neutron capture reaction.}\label{fig_NLD_Eu_1GK}
\end{figure}

In figure \ref{fig_NLD_Eu_1GK} we plot the predicted nuclear level density for neutron captures at a temperature of 1 GK for several neutron-rich isotopes of Eu. This calculation is an example of the overall decreasing trend of the level density in line with the decreasing separation energies. It is also worth noting the generally good agreement of the various models around $A=154$, as opposed to the emergence of significant variations among models above $A=177$.

\begin{figure}[h]
    \includegraphics[width=\columnwidth]{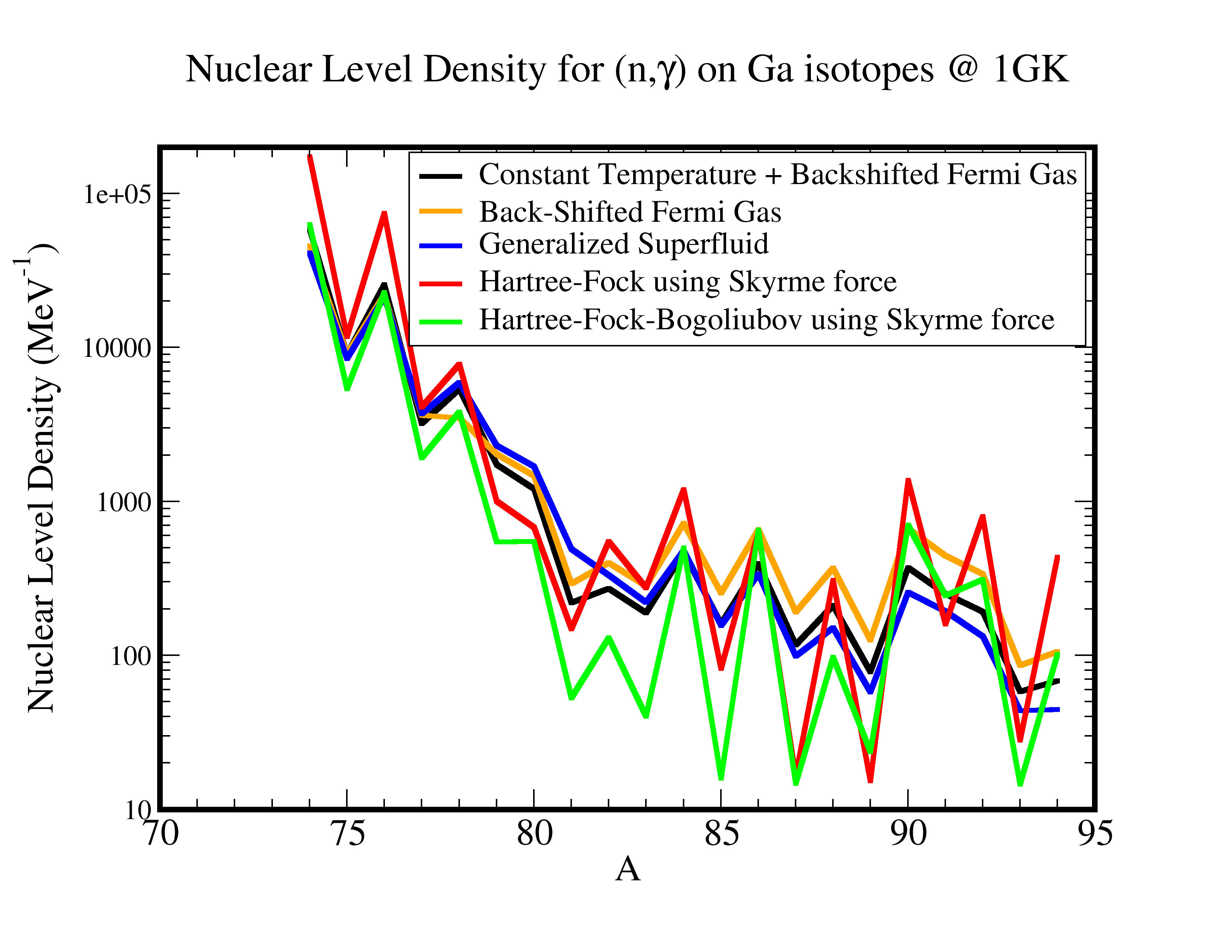}
    \caption{Plot of calculated nuclear level density in neutron capture using various models, for neutron-rich isotopes of gallium. We plot the level density for the compound nucleus created by neutron capture at a stellar temperature of 1~GK. The horizontal axis corresponds to the mass number of the target in the neutron capture reaction.}\label{fig_NLD_Ga_1GK}
\end{figure}

An example of the effect of local nuclear structure and how the model predictions affect the level density description we present in figure \ref{fig_NLD_Ga_1GK} where we calculate the level density for neutron-rich isotopes of gallium at 1GK. The two Hartree-Fock approaches (red and green lines in figure \ref{fig_NLD_Ga_1GK}) provide a \textcolor{red}{level density} rich in local features
that has no counterpart in the three phenomenological models that to a large extent follow the same trend. In particular, the predictions of the Hartree-Fock-Bogoliubov plus combinatorial model (green line in figure \ref{fig_NLD_Ga_1GK}) exhibit sharp odd-even effects that drive the level density much lower than any other model in the region between A=80 and A=87.   
\subsection{Results for E(1) gamma strength functions}\label{subsection_GSF_n_rich}
Gamma ray emission in the Hauser-Feshbach model (see \cite{Bartholomew1973} for a complete description of the underlying theory) is described by the gamma-ray transmission coefficient. For a gamma transition with energy $ E_{\gamma} $ and angular momentum $ l $ it is given by,
\begin{equation}
T_{l}=2\pi f_l(E_{\gamma})E_{\gamma}^{2l+1}
\end{equation}
where $ f_l(E_{\gamma}) $ is the gamma-ray strength and $ E^{2l+1} $ is the energy dependence factor for the particular multipolarity. The gamma-ray strength is proportional to the average level density of the emitting compound $ \rho(E_{\lambda}) $ at an excitation energy E$ _{\lambda} $ and the average radiative width $ \Gamma_{\gamma}(E_\gamma) $ (provided in calculations by the various gamma-ray strength function models).
\begin{equation}
f_{l}(E_{\gamma})=\Gamma_{\gamma}(E_\gamma)\rho({E_\lambda})E_\gamma ^{-(2l+1)}
\end{equation}
The basic underlying assumption is that of the Brink hypothesis. It assumes that any excited level is expected to decay to another level with probability given by the appropriate gamma strength distribution for the multipole corresponding to the angular momentum and parity difference between the two levels.

\begin{figure}[h]
    \includegraphics[width=\columnwidth]{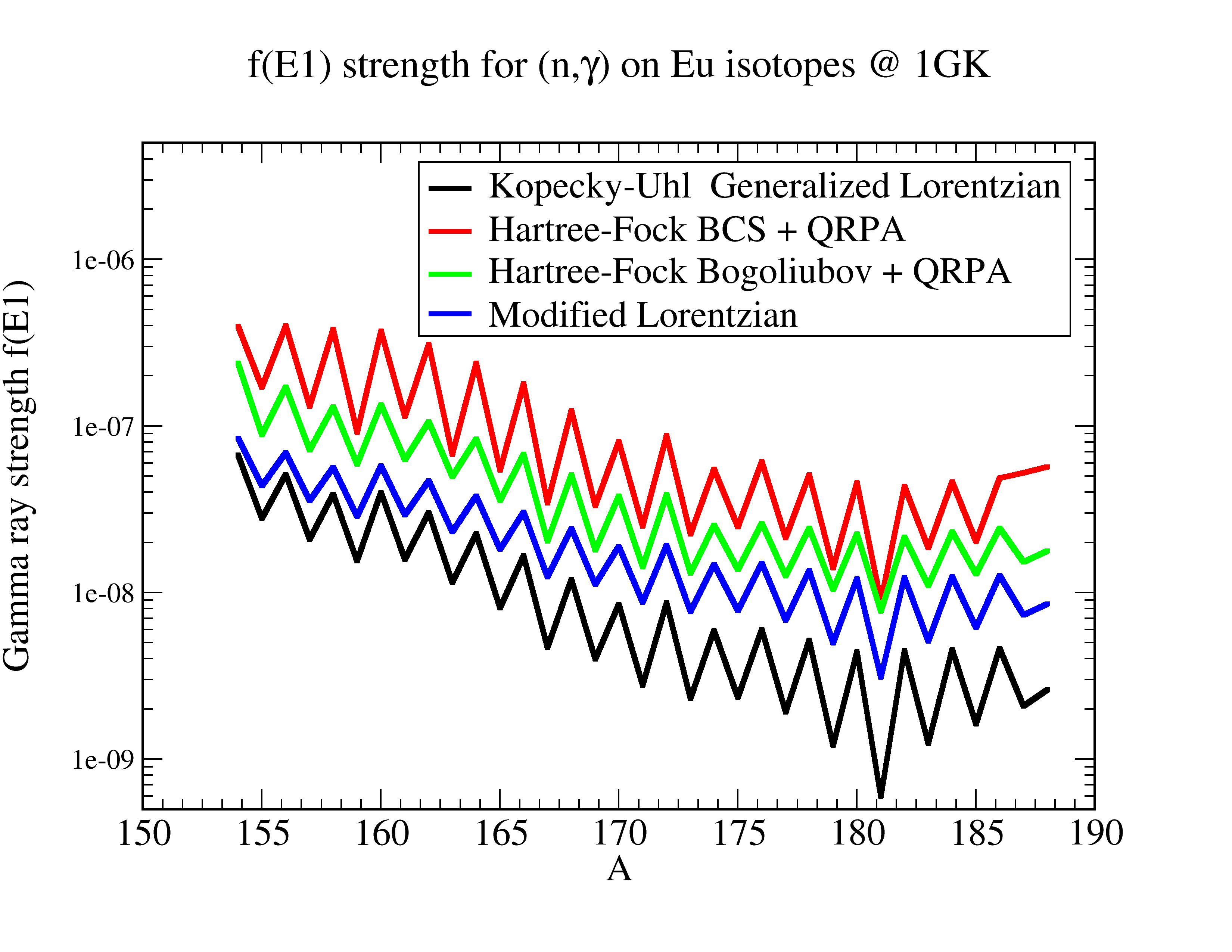}
    \caption{Plot of calculated E(1) strength in neutron capture using various models for neutron-rich isotopes of europium. We plot the strength for the compound nucleus at a stellar temperature of 1~GK. The horizontal axis corresponds to the mass number of the target in the neutron capture reaction.}\label{fig_E1_Eu_1GK}
\end{figure}

\begin{figure}[h]
    \includegraphics[width=\columnwidth]{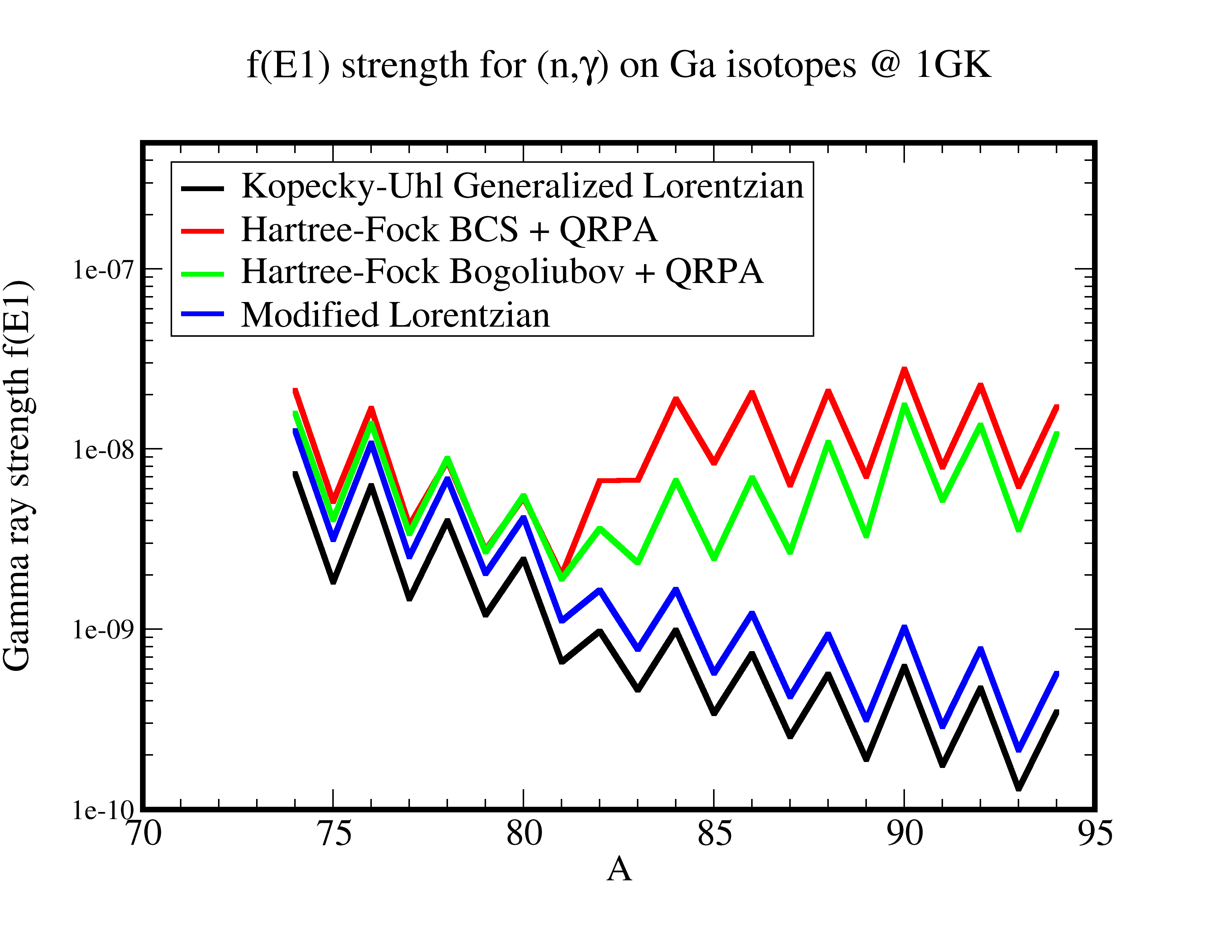}
    \caption{Plot of calculated E(1) strength in neutron capture using various models for neutron-rich isotopes of gallium. We plot the strength for the compound nucleus at a stellar temperature of 1~GK. Note the sharp increase in E(1) strength predicted by the two Hartree-Fock + QRPA for isotopes above A = 81. The horizontal axis corresponds to the mass number of the target in the neutron capture reaction.}\label{fig_E1_Ga_1GK}
\end{figure}

Based on the above model of gamma-ray emission we expect that as we move towards the driplines, the gamma strength of a particular multipole will generally decrease by a factor roughly proportional to the decreasing density of levels provided that the general shape of $ \Gamma_{\gamma}(E_\gamma) $ does not change radically from one isotope to the next.  
In figure \ref{fig_E1_Eu_1GK}, the E(1) gamma-ray strength of neutron-rich europium isotopes at a temperature of 1~GK exhibits a generally decreasing trend. It is interesting to note that the microscopic models produce a significantly larger gamma strength compared to the phenomenological ones while maintaining more or less the same qualitative trend. This observation is a direct consequence of the need to renormalize the phenomenological gamma-ray strengths as mentioned in section \ref{CalculationsHF}. 
In figure \ref{fig_E1_Ga_1GK}, we observe a different behavior of the E(1) strength for the neutron-rich gallium isotopes. Here, the microscopic models predict a less smooth dependence on neutron number with an abrupt increase in the strength for $A \geq $82. The phenomenological models do not reproduce this trend at all. 

\begin{figure}[h]
    \includegraphics[width=\columnwidth]{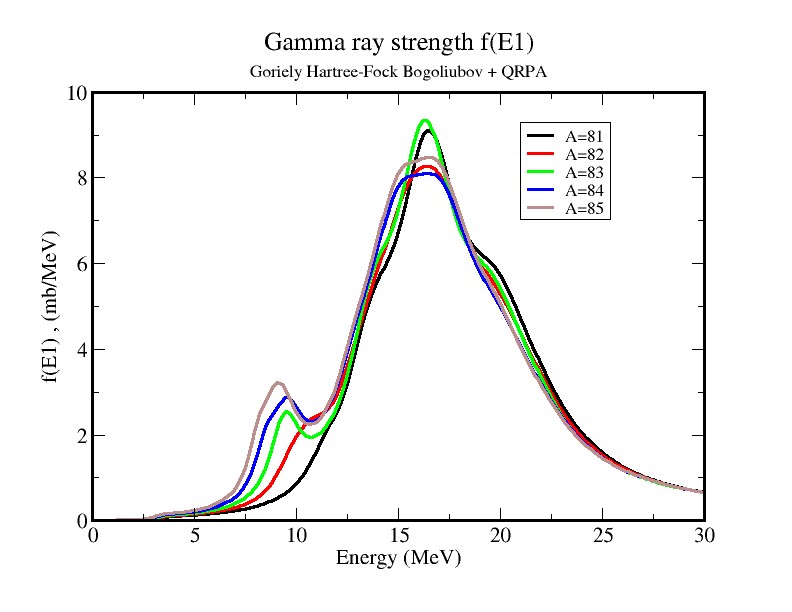}
\caption{Plot of the evolution of E(1) strength as calculated using the Hartree-Fock-Bogoliubov plus QRPA calculation of Goriely et al. Each colored line corresponds to a different neutron-rich isotope of Ga. For $A > $ 81, the calculations predict the appearance of additional strength at a lower gamma energy than typically predicted. Phenomenological models do not reproduce this feature that is a product of the HFB+QRPA calculation.}\label{fig_E1_Ga}
\end{figure}

The explanation for this odd behavior is given in figure \ref{fig_E1_Ga}, where we plot the E(1) gamma strength for gallium isotopes with $A=81$ and above as a function of gamma-ray energy. The predicted strength distribution changes above the $N=50$ shell closure. Additional strength toward the low energy tail of the distribution is expected according to the HFB plus QRPA calculations. This change alters
the gamma-ray strength of the $ ^{82} $Ga compound in the $ ^{81} $Ga(n,$ \gamma $) reaction drastically, increasing the low lying gamma strength on the tail of the E(1) distribution that is accessible at 1~GK.   

\subsection{Effect on reaction rates}\label{subsection_Effect_on_rates}
\begin{figure}[h]
    \begin{subfigure}[h]{\columnwidth}
        \includegraphics[width=0.8\linewidth]{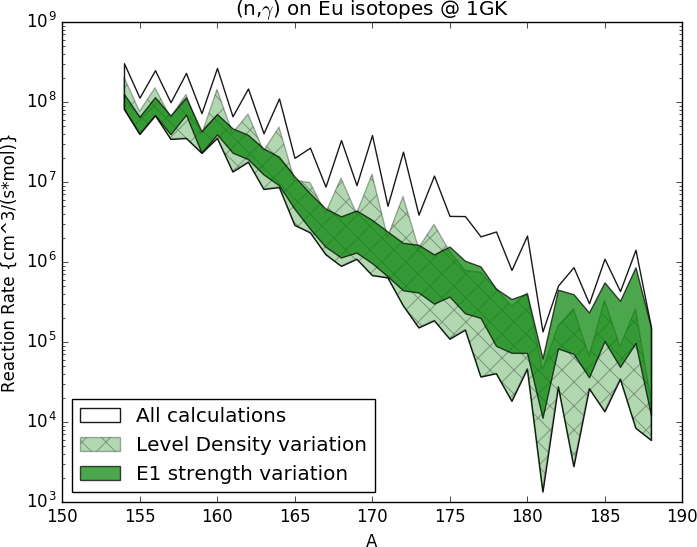}
        \caption{Effect of the variation of the two statistical properties, level density (light green hatch) and gamma ray strength (dark green) to the reaction rate. The white area includes the combined effect of both level densities and gamma strengths to the rate.}
        \label{figure:SpreadGSFLDEu1GK}
    \end{subfigure}

\begin{subfigure}[h]{\columnwidth}
    \includegraphics[width=0.925\linewidth]{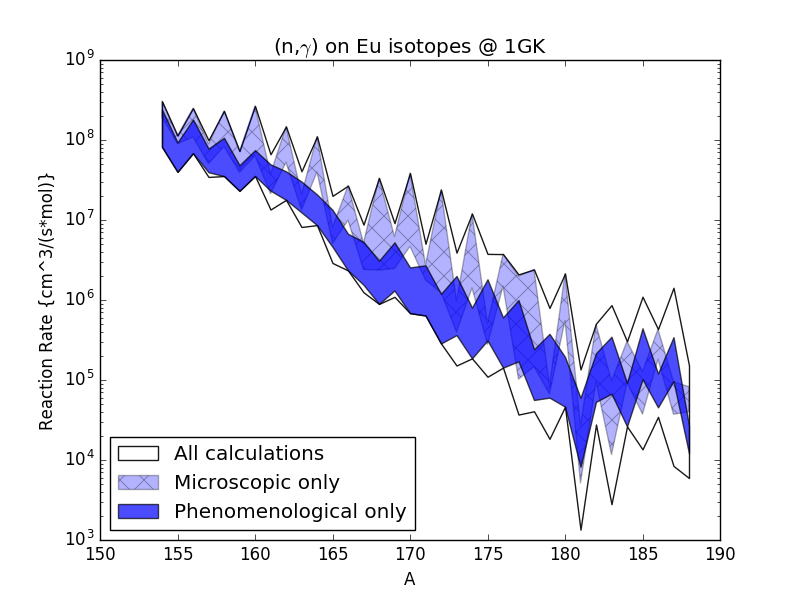}
    \caption{Effect of the choice of microscopic (light blue hatch) or phenomenological (dark blue) models to calculate the two statistical properties, level density and gamma-ray strength to the reaction rate. The white area includes the combined effect of both level densities and gamma strengths to the rate.}
    \label{figure:SpreadMicMacEu1GK}
\end{subfigure}
\caption{Isotopically mapped results of calculations of the neutron capture reaction rate for europium. Different color codings are used to show the effect of model choices used in the calculations for the two statistical properties (see captions under each plot).}
\end{figure}

The above observations regarding the level density and gamma-ray strength have a direct effect on the neutron capture reaction rates that the various theoretical models predict. In figures \ref{figure:SpreadGSFLDEu1GK} and \ref{figure:SpreadMicMacEu1GK} we show the effect of level density and gamma strength variations on neutron capture in europium isotopes for a temperature of 1~GK.  
We observe that for the europium case, the principal cause of rate variation comes from the level density calculations. Above $A=181$, the divergence between various level density calculations increases, driving a corresponding variation in the neutron capture rates. For the odd-odd isotopes of europium below $A=181$ (see figure  \ref{figure:SpreadMicMacEu1GK}), the microscopic models predict larger reaction rates than their odd-even counterparts, causing a large offset between the two groups of calculations.\\
\begin{figure}[h]
    \begin{subfigure}[h]{\columnwidth}
        \includegraphics[width=0.8\linewidth]{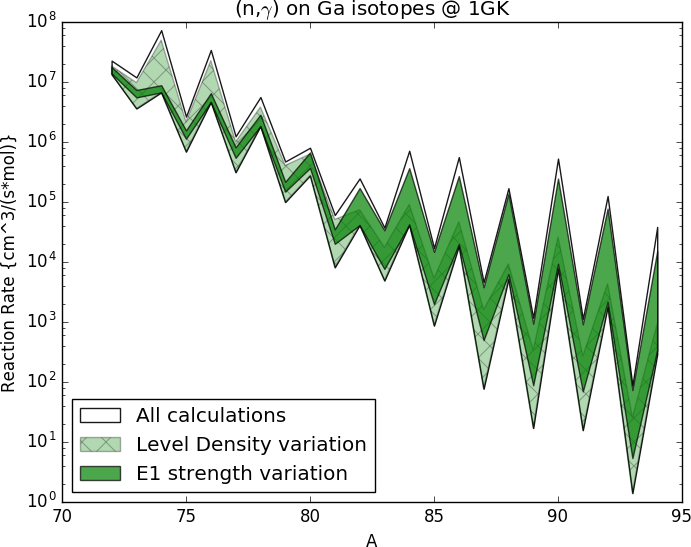}
        \caption{Effect of the variation of the two statistical properties, level density (light green hatch) and gamma ray strength (dark green) to the reaction rate. The white area includes the combined effect of both level densities and gamma strengths to the rate.}
        \label{figure:SpreadGSFLDGa1GK}
    \end{subfigure}

\begin{subfigure}[h]{\columnwidth}
    \includegraphics[width=0.92\linewidth]{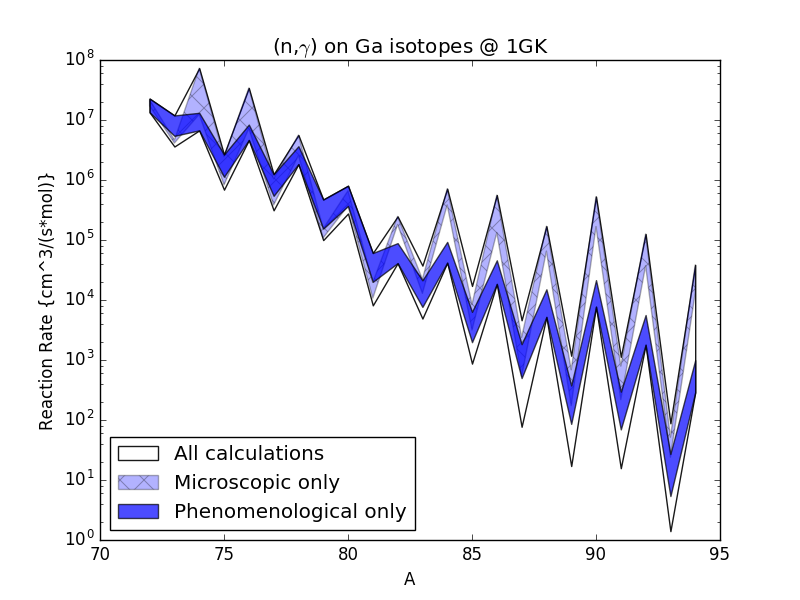}
    \caption{Effect of the choice of microscopic (light blue hatch) or phenomenological (dark blue) models to calculate the two statistical properties, level density, and gamma-ray strength to the reaction rate. The white area includes the combined effect of both level densities and gamma strengths to the rate.}
    \label{figure:SpreadMicMacGa1GK}
\end{subfigure}
\caption{Isotopically mapped results of calculations of the neutron capture reaction rate for gallium. Different color codings are used to show the effect of model choices used in the calculations for the two statistical properties (see captions under each plot)}
\end{figure}
The results are qualitatively different for Ga isotopes in figures \ref{figure:SpreadGSFLDGa1GK} and \ref{figure:SpreadMicMacGa1GK}. Here, the gamma strength dominates the reaction rate uncertainty. In accordance with the results of figure \ref{fig_E1_Ga_1GK} the disagreement between the various reaction rate calculations increases drastically above $A=81$. As the shell closure at $N=50$ is crossed, the microscopic gamma strength models predict extra low-lying strength for the more neutron-rich gallium isotopes. The phenomenological models produce a smoother gamma strength variation as a function of neutron number even across the closed neutron shell causing a marked offset between microscopic and phenomenological reaction rates.

\subsection{A closer look at two special cases of neutron capture rates.}\label{subsection_special}
It is worthwhile to study some individual cases along these two isotopic lines to appreciate how the dependence of statistical properties on the excitation energy affects the calculated reaction rates as a function of temperature. Based on earlier sensitivity studies by Mumpower and Surman \cite{mumpower2016impact}, \cite{Mum12a},\cite{Sur14a} we present here calculation results for neutron capture on the nuclei $ ^{165} $Eu, and $ ^{81} $Ga, both identified for the sensitivity of r-process nucleosynthesis yields to their neutron capture rate.
\subsubsection{Results for the $^{165}$Eu(n,$ \gamma $)$ ^{166} $Eu reaction rate.}
In a sequence of sensitivity studies performed in the last decade, the reaction rate of the $ ^{165} $Eu(n,$ \gamma $)$ ^{166} $Eu reaction appears to affect the abundance yields of r-process nucleosynthesis for a variety of potential astrophysical scenarios. Notably, this reaction rate has a significant effect on the formation of the rare earth peak, a feature of the r-process abundance pattern that is considered a sensitive benchmark in comparisons of calculations with observed abundances \cite{mumpower2016impact}, \cite{Mum12a}.

\begin{figure}[h]
    \begin{subfigure}[h]{\columnwidth}
        \includegraphics[width=0.8\linewidth]{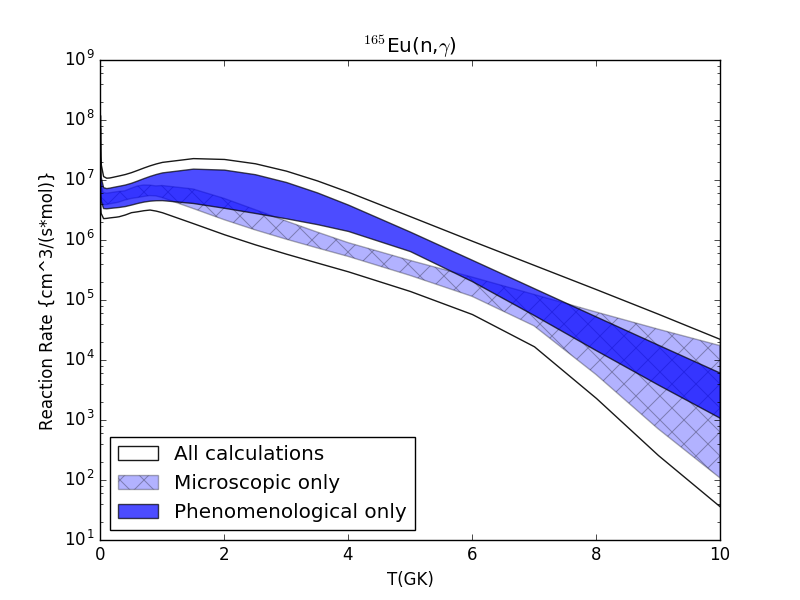}
        \caption{Range of reaction rate results calculated by microscopic only models (light blue hatch), and by phenomenological only models (dark blue).}
        \label{figure:Spread165Eu}
    \end{subfigure}
    
    \begin{subfigure}[h]{\columnwidth}
        \includegraphics[width=0.8\linewidth]{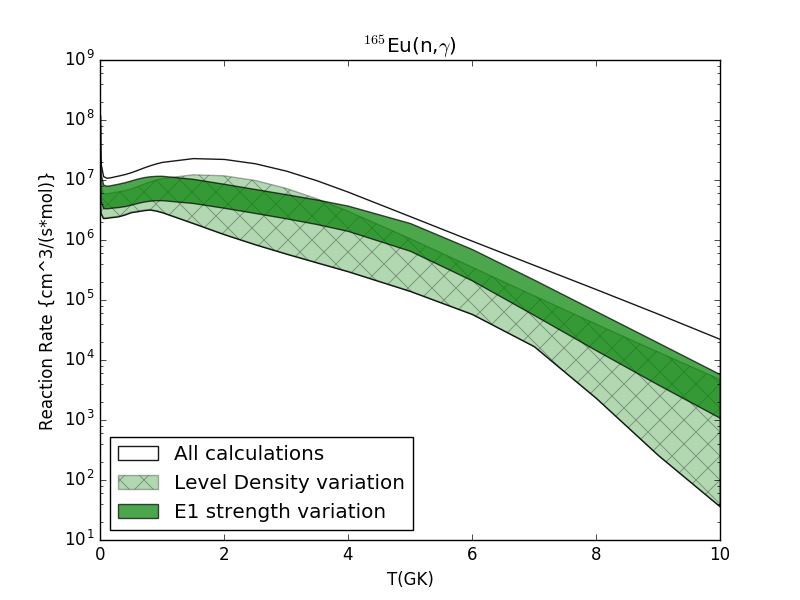}
        \caption{Range of reaction rate results obtained with the level density models of table \ref{table:NLD_GSF_models} with the gamma strength model fixed (light green hatch), and vice versa by varying the gamma strength description with the level density model fixed (dark green).}
        \label{figure:SpreadGSFLD165Eu}
    \end{subfigure}
    \caption{Range of calculated reaction rates as a function of stellar temperature for neutron capture on $ ^{165} $Eu(n,$ \gamma $). The results are color-coded to show various contributions to the calculated reaction rates. In both top and bottom, the contribution from all combinations of microscopic and phenomenological models on table \ref{table:NLD_GSF_models} corresponds to the white region.}
\end{figure}

Our study, summarized in figures \ref{figure:Spread165Eu} and \ref{figure:SpreadGSFLD165Eu}, shows that it is reasonable to assume an uncertainty of roughly one order of magnitude in the temperature range of 1-3~GK for this rate. The model parameters responsible for this uncertainty vary with the stellar environment temperature. Under 1~GK most of the uncertainty comes from the description of the gamma decay of the $ ^{166} $Eu compound nucleus. Above 1~GK, the level density starts to become more important and finally dominates the rate uncertainty. 
\subsubsection{Results for the $ ^{81} $Ga(n,$ \gamma $)$ ^{82} $Ga reaction rate.}
The reaction rate of the $ ^{81} $Ga(n,$\gamma  $)$ ^{82} $Ga reaction can influence the abundance pattern predicted for a range of different characteristic weak r-process trajectories in a set of more than fifty sensitivity studies reported by Surman et al. \cite{Sur14a}. Moreover, the value of this reaction rate caused the largest variations in the final abundances of the weak r-process, even when the results of all fifty sensitivity studies were accumulated. $ ^{81} $Ga is a special nucleus from the point of view of the nuclear structure as it is situated on the $N=50$ shell closure and is in the path of the weak r-process. While the aforementioned sensitivity studies varied reaction rates in the weak r-process by a factor of 100, the present work can provide a more informed estimate of the expected uncertainty of this reaction rate.

\begin{figure}
    \begin{subfigure}[h]{\columnwidth}
        \includegraphics[width=0.8\linewidth]{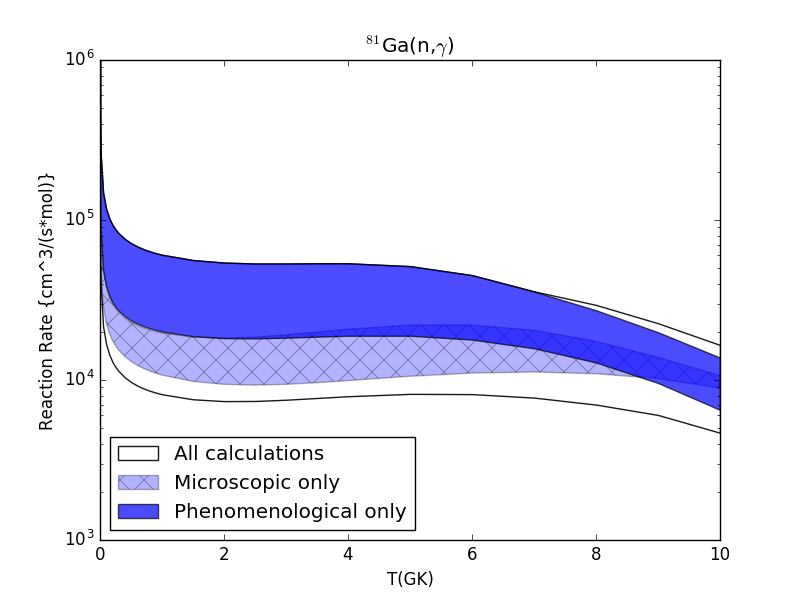}
        \caption{Range of reaction rate results calculated by microscopic only models (light blue hatch), and by phenomenological only models (dark blue).}
        \label{figure:Spread81Ga}
    \end{subfigure}
    
    \begin{subfigure}[h]{0.45\textwidth}
        \includegraphics[width=0.8\linewidth]{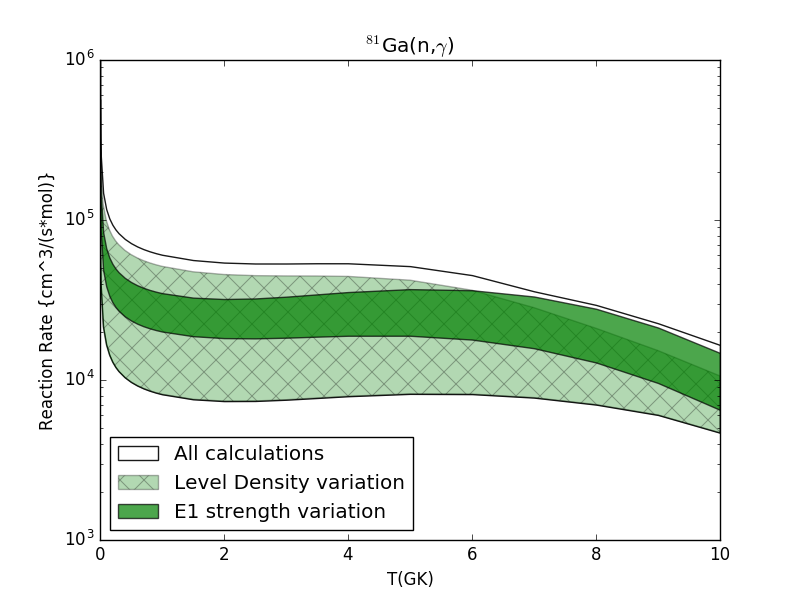}
        \caption{Range of reaction rate results obtained with the level density models of table \ref{table:NLD_GSF_models} with the gamma strength model fixed (light green hatch), and vice versa by varying the gamma strength description with the level density model fixed (dark green).}
        \label{figure:SpreadGSFLD81Ga}
    \end{subfigure}
    \caption{Range of calculated reaction rates as a function of stellar temperature for neutron capture on $ ^{81} $Ga(n,$ \gamma $). The results are color-coded to show various contributions to the calculated reaction rates. In both top and bottom, the contribution from all models on table \ref{table:NLD_GSF_models} corresponds to the white region.}
    \label{figure:81Ga}
\end{figure}

Based on the results of our Hauser-Feshbach calculations plotted in figures \ref{figure:SpreadGSFLDGa1GK} and \ref{figure:SpreadMicMacGa1GK} , an uncertainty factor of 100 is reasonable for some of the more neutron-rich isotopes of gallium. In the case of the $ ^{81} $Ga reaction rate, the results of figure \ref{figure:81Ga} suggest that we can reasonably expect an uncertainty of one order of magnitude. This uncertainty is dominated by the modeling of the level density. It has to be noted though, that the results of figures \ref{figure:SpreadGSFLDGa1GK} and \ref{figure:SpreadMicMacGa1GK} suggest that the gamma strength becomes a dominant contributor to the uncertainty for more neutron-rich gallium isotopes.

\subsubsection{Summary of results for the whole neutron-rich region}
\begin{figure*}[h]
\includegraphics[width=\textwidth]{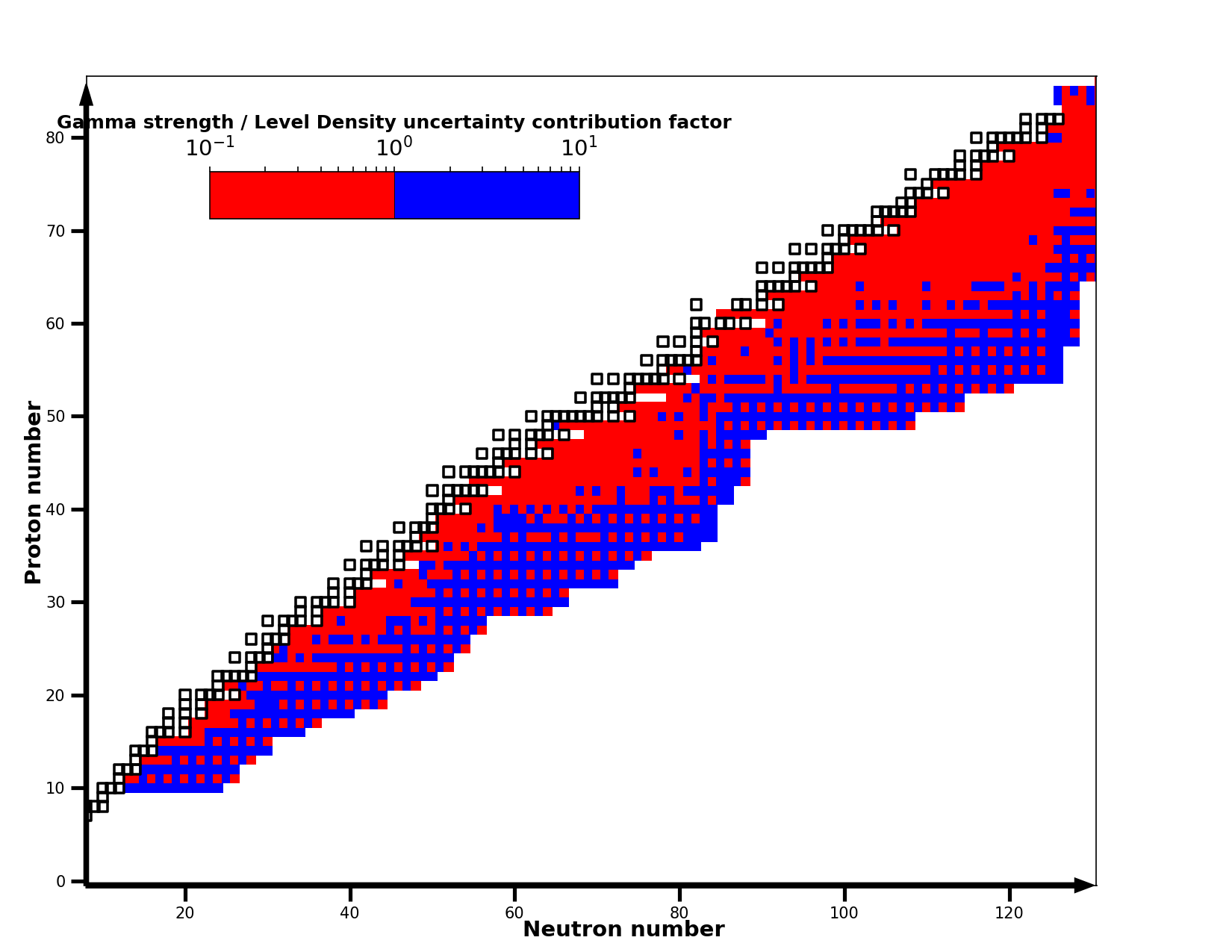}
\caption{The figure shows on the table of isotopes, whether the Gamma strength or level density model dominates in the contribution to the variation of the reaction rate. In the color bar values greater than 1 (blue) correspond to a domination of the Gamma Strength, while values less than 1 (red) show domination of Level density modeling}\label{figure:Contributor_1GK}
\end{figure*}

One last question is whether the level density or the gamma-ray strength modeling is responsible for the extrapolation uncertainty in the various regions of the nuclear chart. In figure \ref{figure:Contributor_1GK} we map on the neutron-rich part of the nuclear chart, the modeling of which statistical property dominates in generating large variations of the reaction rate. The general observation is that the rates are more sensitive to extrapolations of level density near stability and more sensitive to the gamma-ray strength toward the dripline. However, this conclusion is not universal and seems to break down for isotopes in two broad regions of the nuclear chart. The first region starts with the neutron-rich nuclei between Zr and Cs, i.e. nuclei with atomic numbers larger than 40 and below the N=82 shell. The second trend-breaking region starts above Eu with neutron-rich isotopes with atomic number 64 and continues partially until the N=126 shell is filled. This last observation suggests that the boundaries of these two regions are areas suitable to focus on future research attempts. Spectroscopic data and direct inferences of the statistical properties in those regions could potentially inform statistical nuclear property modeling away from stability. 

\subsection{Monte-Carlo propagation to nucleosynthesis}\label{subsection_MC_nucleosynthesis}

\begin{figure}
    \begin{subfigure}[h]{\columnwidth}
        \includegraphics[width=\linewidth]{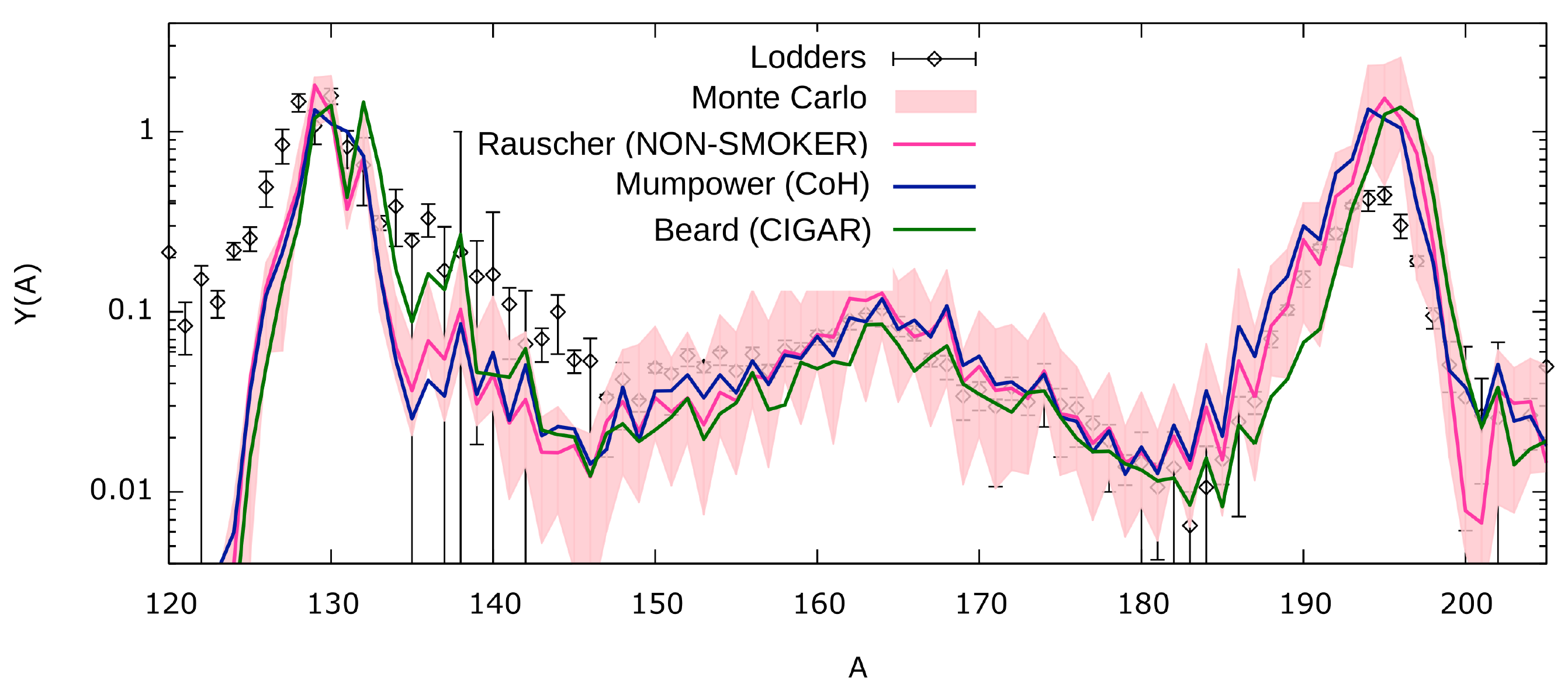}
        \caption{Results for a low-entropy hot neutrino driven wind environment.}
        \label{figure:MC_low_entropy_hot}
    \end{subfigure}
    
    \begin{subfigure}[h]{\columnwidth}
        \includegraphics[width=\linewidth]{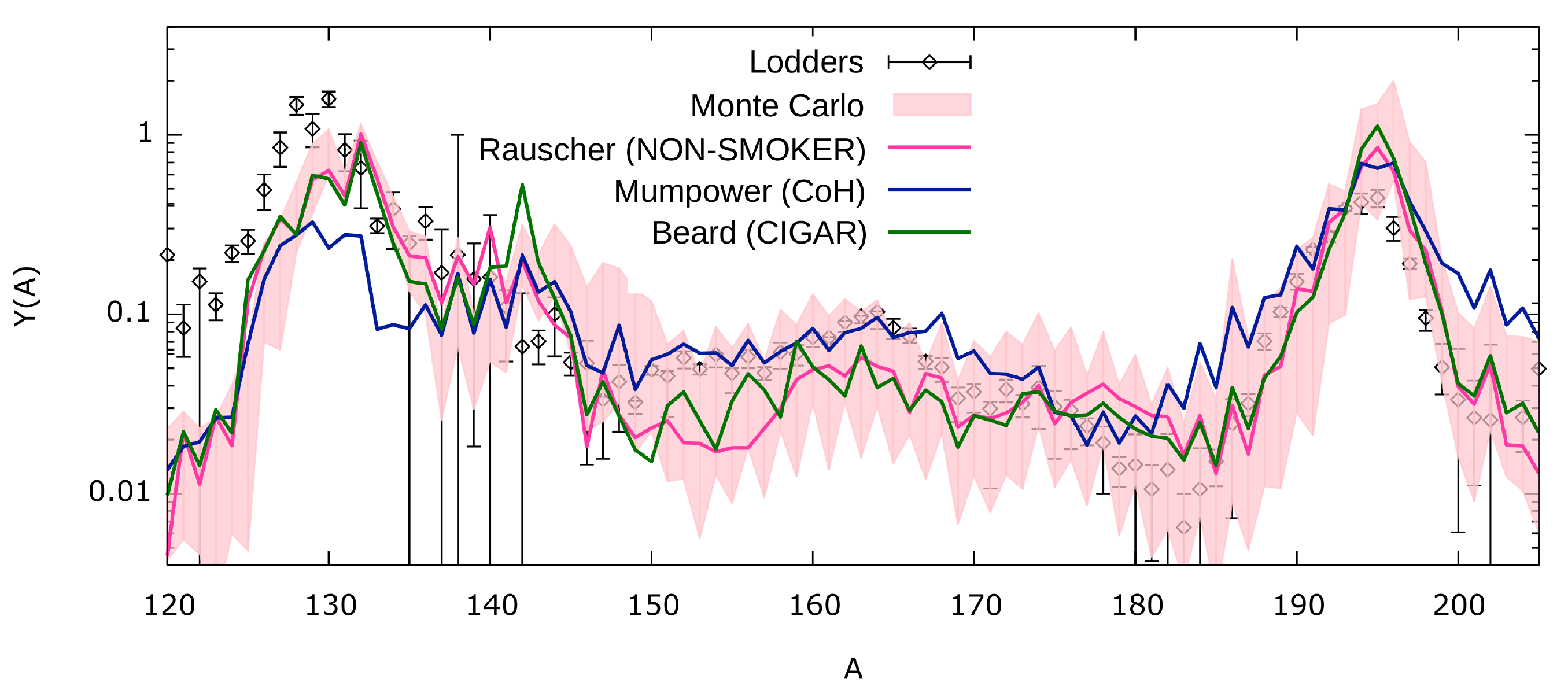}
        \caption{Results for a neutron star merger environment.}
        \label{figure:MC_neutron_star}
    \end{subfigure}
    \caption{Monte Carlo study of the effect of the reaction rate uncertainties identified in this work for two nucleosynthesis scenarios. The study is compared with single network calculations using specific neutron capture rates. Abundances are plotted as a function of mass number. Pink area: Monte-Carlo. Red line: single network with reaction rates from Rauscher et al \cite{Rau01a}. Blue line: idem, by Mumpower et al \cite{Mumpower2018}, Green line: idem,with rates by Beard et al \cite{Beard:2015cra}. Circles: Normalized r-process abundances based on \cite{Lodders2010}}
    \label{fig:MonteCarlo}
\end{figure}

The uncertainties inherent in the reaction rate calculations described above were propagated to nucleosynthesis in a Monte Carlo study, as described in section \ref{MonteCarlo}. We show the results for two characteristic environments, that of a low-entropy wind such as may occur in outflows from neutron star merger (NSM) accretion disks, similar to those in, e.g., \cite{Just+15}, and that of the dynamical ejecta from a neutron-star merger as in \cite{Mumpower+18} (figures \ref{figure:MC_low_entropy_hot} and \ref{figure:MC_neutron_star} respectively). The low-entropy wind is characterized by an extended ($n$,$\gamma$)-($\gamma$,$n$) equilibrium such that the influence of neutron capture rates on the predicted abundances is limited to the late-time decay toward stability. In the NSM dynamical ejecta conditions adopted here, on the other hand, equilibrium fails early, and the r-process reaction flow proceeds close to the neutron drip line. As a result, nuclei farther from stability are accessed, where the neutron capture rate variations are largest, and their rates are more impactful since the species are populated out of equilibrium. Therefore, the range of results of the Monte Carlo (pink band in figure \ref{fig:MonteCarlo}) is broader for the NSM dynamical ejecta conditions than for the hot wind conditions.

The results of the Monte Carlo study are compared against single calculations of the same network. For each of the non-Monte Carlo calculations, the neutron capture rates are provided from a single Hauser-Feshbach calculation each time, with a single choice of level density model, strength function model, etc. i.e., the neutron capture rates are not randomly sampled from the results of multiple calculations. Single calculations are important as benchmarks to ensure that the random sampling of reaction rate values does not create spurious or unphysical features. For completeness, we note the code used to generate each of these reaction rate sets in the figure legend, although this is meant as a comparison of models and not of software packages.

The range of results of the Monte Carlo (pink band in figure \ref{fig:MonteCarlo} shows more variation than the three regular network calculations (solid lines). This increased variation is expected since the three calculations do not necessarily span the range of Hauser Feshbach parametrizations that the Monte Carlo calculation includes. However, it is notable that the Monte Carlo abundances vary similarly to the three single reaction rate calculations in a few mass regions, while they are surpassed in variation by the spread of single calculations in one mass region. 
For example, a similar magnitude of abundance variation is observed in the case of the low-entropy wind of figure \ref{figure:MC_low_entropy_hot} for isotopes with $130 <A<140$, $180<A<195$, and $A\approx200$. 
The non-Monte Carlo results diverge even more than the Monte Carlo study in the NSM dynamical ejecta scenario of figure \ref{figure:MC_neutron_star} for nuclei with $125<A<135$. These observations serve as a suggestion that the Monte Carlo technique does not globally overestimate the nucleosynthesis yield variation. The Monte Carlo results are within the range of what could be obtained by performing traditional network calculations using theoretical neutron capture rates, and its use as a tool to explore the sensitivity of abundance yields to the model uncertainties inherent in Hauser Feshbach extrapolations seems justified.

The Monte Carlo calculations of figure \ref{fig:MonteCarlo} can be compared with the study of figure \ref{fig:R-process_Impact2} in the Introduction. The pink band of results suggests an abundance uncertainty for most isotopes that is comparable to the results obtained by randomly varying each reaction rate within a factor of 10 uncertainty in the sensitivity study of figure \ref{fig:R-process_Impact2}. Within the uncertainty band, the nucleosynthesis calculation for both astrophysical scenarios generally agrees with the shape of the r-process abundances pattern for $145<A<190$. However, the magnitude of the uncertainty does not allow to extract any conclusion regarding the detailed shape of the calculated abundances in the region of agreement.
\section{Conclusion} \label{Conclusion}
Efforts to solve the puzzle of the synthesis of elements heavier than iron depend critically on the micro-physics input to astrophysics models. Ideally, a reliable set of experimentally measured neutron capture rates for most of the nuclei involved in the r-process is required. Due to the technological limitations that prevent us from developing a reaction target made out of neutrons or some other equivalent accelerator apparatus, we can not currently use the available radioactive beams to measure neutron capture reactions on short-lived nuclei directly. Hence, neutron capture rates for r-process currently come from theoretical calculations that contain a large number of parameters that are not adequately constrained. It is the consensus of the community that these calculations infer large uncertainties to astrophysics calculations. 

To evaluate the yield outcome of various astrophysics scenarios we need to be able to reproduce in nucleosynthesis calculations, complex features of abundance yield patterns. For such comparisons to be meaningful, uncertainties in the nuclear input that affect nucleosynthesis calculations have to be identified, and their influence evaluated. To address this need, we investigated the sources of uncertainty that are most influential to the extrapolation of Hauser-Feshbach calculations away from stability and traced them back to the description of model ingredients that mostly influence neutron capture reaction rates, namely the level density, and the gamma-ray strength distribution. We calculated reaction rates using a number of adequate level density and gamma strength models for the neutron-rich isotopes of elements from oxygen to uranium. For this extensive list of isotopes, we compared the results of different calculations for each reaction rate and calculated the ratio of minimum to maximum result for temperatures up to 10GK. We found results that vary up to a few orders of magnitude for each reaction rate and studied how the combined effect of inconsistent model predictions for the level density and the $\gamma$-ray strength created increased uncertainty and reduced the reliability of neutron capture rates away from stability. Based on these results it is clear that improvements in the current reaction theory and in particular the development of better microscopic models for gamma strengths and level densities are imperative for as long as we rely on Hauser-Feshbach theory to calculate neutron capture rates.
\begin{acknowledgments}
We wish to acknowledge support from the National Science Foundation under Grant No. PHY-1430152 (JINA Center for the Evolution of the Elements), the College of Science and Engineering of Central Michigan University and conversations during the CETUP* workshop. 
M.M. was supported by the US Department of Energy through the Los Alamos National Laboratory. Los Alamos National Laboratory is operated by Triad National Security, LLC, for the National Nuclear Security Administration of U.S.\ Department of Energy (Contract No.\ 89233218CNA000001). 
M.M. further acknowledges support from the Laboratory Directed Research and Development program of Los Alamos National Laboratory under project number 20190021DR. 
\end{acknowledgments}
%\input{extras}

%\nocite{*}
\bibliography{HFSens.bib}% Produces the bibliography via BibTeX.
\end{document}